\documentclass[12pt]{iopart}

\usepackage{iopams}
\usepackage{graphicx}

\usepackage{amssymb}
\usepackage{amsmath}

\usepackage{mathbbol}

\usepackage[colorlinks, linkcolor=blue, citecolor = blue, urlcolor = blue]{hyperref}
\usepackage{cite}

\usepackage[utf8]{inputenc}
\usepackage[T1]{fontenc}
\usepackage[]{units}
\usepackage[english]{babel}
\usepackage{tabularx}

\newcommand{\dif}{\mathrm{d}}%
\newcommand{\Sp}{{\mathcal{S}}}%
\newcommand{\Sphere}{{\mathbb{S}_{2}}}%

\newcommand{\rb}{\mathbf{r}}%
\newcommand{\pb}{\mathbf{p}}%
\newcommand{\pbc}{\hat{p}}%
\newcommand{\qb}{\mathbf{q}}%
\newcommand{\qbc}{\hat{q}}%
\newcommand{\uu}{\mathbf{\hat{u}}}%
\newcommand{\norm}[1]{\lVert#1\rVert}%
\newcommand{\C}{\mathbb{C}}%

\newcommand{\Nsh}{n}%
\newcommand{\Nth}{N_{\theta}}%
\newcommand{\Nphi}{N_{\phi}}%
\newcommand{\TempR}{{\varepsilon}}%

\newcommand{\Tang}[2]{{\mathrm{T}_{#1}{#2}}}%
\newcommand{\TS}{{\Tang{\rb}{\Sp}}}%

\makeatletter
\newcommand*\bigcdot{\mathpalette\bigcdot@{.5}}
\newcommand*\bigcdot@[2]{\mathbin{\vcenter{\hbox{\scalebox{#2}{$\m@th#1\bullet$}}}}}
\makeatother

\newcommand{\GradS}{{\operatorname{grad}_{\Sp}}}%
\newcommand{\rotS}{{\operatorname{rot}_{\Sp}}}%
\newcommand{\RotS}{{\operatorname{Rot}_{\Sp}}}%
\newcommand{\DivS}{{\operatorname{div}_{\Sp}}}%
\newcommand{\Laplace}{\boldsymbol{\triangle}}%
\newcommand{\LaplaceS}{\Laplace_{\Sp}}%
\newcommand{\LaplaceDR}{\Laplace_{\mathrm{dR}}}%

\newcolumntype{L}{>{$}l<{$}}%
\newcolumntype{R}{>{$}r<{$}}%
\newcolumntype{Y}{>{\centering\arraybackslash}X}%

\newcommand{\density}{\psi}
\newcommand{\activity}{v_0}

\newcommand{\defectVelo}{v}

\newcommand{\numDefect}{N}

\newcommand{\surf}{\mathcal{S}}
\newcommand{\normalSurfC}{\nu}
\newcommand{\normalSurf}{\boldsymbol{\normalSurfC}}
\newcommand{\ProjSurf}{\boldsymbol{\Pi}}
\newcommand{\xb}{\mathbf{x}}

\newcommand{\dirNematic}{\mathbf{d}}

\newcommand{\qtensorC}{Q}
\newcommand{\qtensor}{\mathbf{\qtensorC}}
\newcommand{\energyDensity}{f}

\newcommand{\wrt}{w.r.t}
\newcommand{\eg}{e.g.}

\begin{document}

\title{Active smectics on a sphere}

\author{Michael Nestler$^1$, Simon Praetorius$^1$, Zhi-Feng Huang$^2$, Hartmut L\"owen$^3$, Axel Voigt$^{1,4}$}

\address{$^1$ Institute of Scientific Computing, Technische Universität Dresden, 01062 Dresden, Germany \\ $^2$ Department of Physics and Astronomy, Wayne State University, Detroit, Michigan 48201, USA \\ $^3$ Institut für Theoretische Physik II: Weiche Materie, Heinrich-Heine-Universität Düsseldorf, 40225 Düsseldorf, Germany \\ $^4$ Center for Systems Biology Dresden, Pfotenhauerstr. 108, 01307 Dresden, Germany}
\ead{axel.voigt@tu-dresden.de}


\begin{abstract}
The dynamics of active smectic liquid crystals confined on a spherical surface is explored through an active phase field crystal model. Starting from an initially randomly perturbed isotropic phase, several types of topological defects are spontaneously formed, and then annihilate during a coarsening process until a steady state is achieved. The coarsening process is highly complex involving several scaling laws of defect densities as a function of time where different dynamical exponents can be identified. In general the exponent for the final stage towards the steady state is significantly larger than that in the passive and in the planar case, i.e., the coarsening is getting accelerated both by activity and by the topological and geometrical properties of the sphere. A defect type characteristic for this active system is a rotating spiral of evolving smectic layering lines. On a sphere this defect type also determines the steady state. Our results can in principle be confirmed by dense systems of synthetic or biological active particles.
\end{abstract}

\section{Introduction}

Liquid crystals consist of particles that can possess both orientational and positional degrees of freedom \cite{de_Gennes_book,Chaikin_Lubensky}. This induces a wealth of ``mesophases'' with partial orientational and positional ordering. In general these phases contain topological defects. For pure positional ordering such defects are characterized by deviations from the equilibrium number of neighbors, and for orientational order alone a director field can be defined and topological defects emerge where the orientation of this field shows a singularity. In smectic phases, orientational and positional orders are present simultaneously and compete with each other, and these defects also interact. Describing and predicting the coarsening of defected, multidomain smectic configurations is a long-lasting problem. Similar to grain growth in crystalline materials, self-similarity in domain coarsening has been found, which provides a scaling law  $t^\alpha$ for the characteristic domain size with time $t$ and a scaling exponent $\alpha$. The exponent $\alpha$ varies between $1/5$ and $1/2$ according to different theoretical \cite{EVG_PRL_1992,EVG_PRA_1992,CM_PRL_1995,BV_PRE_2001,BV_PRE_2002,AR_NJP_2008} and experimental \cite{Hetal_S_2000,PD_PRL_2001} investigations in 2D space. The coarsening process is usually dominated by the motion of grain boundaries, which move over large distances and are driven by the curvature of the smectic lines and internal distortions within domains. These distortions are positional defects, in particular disclinations and dislocations. The motion of dislocations \cite{PP_JAP_1975} and the motion of disclinations \cite{Hetal_PRE_2002,AR_NJP_2008} in smectics are coupled by topological constraints and their interactions influence the coarsening process. This complexity might explain the broad spectrum of reported values of scaling exponent. Results on power law scaling for defect densities in 2D space are more narrow, for which the disclination density scales as $t^{-1/2}$ and the dislocation density scales as $t^{-1/3}$ (see \cite{Hetal_S_2000,Hetal_PRE_2002,BV_PRE_2002,AR_NJP_2008}).

Here we address two questions: How is this coarsening process influenced if the smectic phase forms on the surface of a sphere, with the director constrained to be tangential, and how does it change if the smectic phase is active? The first introduces topological constraints resulting from the Poincar\'e-Hopf theorem and Euler’s theorem on polyhedra, and forces the defects (orientational and positional) to be present also in equilibrium. The second brings the system into nonequilibrium and there is no relaxation towards equilibrium but rather towards a nonequilibrium steady state. Smectics on spherical surfaces, the so-called smectic shells, have been experimentally realized \cite{lopez2011nematic,liang2013tuning}. The topological and geometrical frustration in these experiments indeed leads to the formation of defects. However, in these experiments the thickness of the smectic layer and the proposed anchoring conditions for the director also influence the equilibrium state. Our study here corresponds to the vanishing thickness limit of an ideal smectic sphere. This case is characterized by two $+1$ (orientational) defects located on opposing poles of the sphere and a director field aligned with geodesic lines connecting these defects. Activity in smectics can be realized by Janus rods which are, for instance, catalytically driven and thus convert chemical energy into mechanical work \cite{Bechingeretal_RMP_2016}. Realizations of active smectics are mostly considered in 2D flat space \cite{Adhyapak_PRL_2013,Chen_PRL_2013}. These systems can be characterized by aligning interaction \cite{Romanczuk_NJP_2016}, nonreciprocality \cite{Saha_PRX_2020}, or nonlinear mutual feedback \cite{Tarama_PRE_2019}. Due to the nontrivial coupling between orientational and positional degrees of freedom with the active driving force, active smectic systems and their defect dynamics are more complex than other active liquid crystals, e.g., active nematics \cite{Doostmohammadi_NC_2018}. To the best of our knowledge, studies of defects in active smectics if constrained to the surface of a sphere, are still lacking. Potential applications for active smectics on surfaces are found in biology. As mentioned in \cite{JPT_PRE_2022}, developing tissues on substrates or myosin filaments in cellular cortex are important examples. The investigation of such systems on a sphere is the first step to understand the complex interplay of orientational and positional defects in multidomain smectic systems under topological constraints, geometric effects, and active driving. 

We study this complex interplay computationally using an active phase field crystal (PFC) model on a sphere \cite{praetorius2018active}. In the following we first review the known results for equilibrium configurations on a sphere and summarize results for the dynamics of the corresponding active systems. We then introduce the active PFC model which has recently been used to model active smectics in 2D space \cite{HLV_CP_2022}, reformulate it on surfaces and simulate the coarsening process on a sphere. While essential processes during coarsening are explained on highly resolved simulations, results on scaling behavior are addressed by large scale statistical investigations. To unveil the influence of the spherical topology and activity, we compare the results with corresponding simulations in flat 2D space and vary the strength of activity.

\section{Review of known results}

\subsection{Equilibrium configurations on a sphere}
Equilibrium configurations for pure orientational or pure positional ordering on a sphere have been intensely studied (see, e.g., \cite{Bowick} for a review). To identify the optimal distribution of interacting particles on a sphere is a classical problem \cite{Thomson_PM_1904,ErberHockney_JPA_1991,Smale_MI_1998}, for which above a certain number of particles there exist non-trivial configurations of defects, the so-called grain boundary scars \cite{BoNeTr00,Bauschetal_Science_2003,BackofenVoigtWitkowski_PRE_2010}, all consistent with Euler’s theorem on polyhedrons. For anisotropic particles the situation becomes more complicated. For orientational ordering \cite{Lubenskyetal_JPII_1992,TVN_RMP_2010,Dzubiella,Zannoni,Shin_PRL_2008,Dhakal_2012,Koning,Napoli,HMS_PRE_2018,Nitschkeetal_PRSA_2018,Nitschkeetal_PRSA_2020}, typically four $+1/2$ defects arrange in a tetrahedral configuration, maximizing their distance from each other. Much less is known for structures that are both orientationally and positionally ordered, such as the smectic phase in curved space. Clearly the simultaneous presence and competition of both local orientational and positional orders put this situation into a much more complex category. For results on a sphere we refer to \cite{Shin_PRL_2008,smallenburg2016close,Kamien,Xing,AVL_SM_2017}. Particle-resolved computer simulations of hard rods indicate a wealth of different states, including equatorial smectics with two $+1$ defects at the poles, consistent with the finding in \cite{lopez2011nematic,liang2013tuning}, and the configurations with four $+1/2$ or six $+1/2$ and two $-1/2$ defects. In these situations the strain at the poles is relieved by separating closely positioned half-strength defects. All these configurations obviously fulfill the topological constraint of the Poincar\'e-Hopf theorem.
 
\subsection{Active liquid crystals on a sphere}

The active liquid crystalline phases, which have been of tremendous interest in recent years \cite{Marchettietal_RMP_2013,Bechingeretal_RMP_2016,BerryRPP18,BowickPRX22,S_NRP_2022}, exhibit qualitatively different behaviors as compared to those in the corresponding passive systems. Their governing dynamics is non-relaxational, as in the far-from-equilibrium pattern formation processes \cite{CrossRMP93}. This type of system on curved surfaces has been termed topological active matter \cite{keber2014topology} as it combines active liquid crystals with topological constraints. Again, studies are mainly concerned with either pure orientational order, i.e., active nematics \cite{keber2014topology,AKV_SR_2017,Apaza_SM_2018,Pearce_PRL_2019,NV_CiCP_2022} showing persistent oscillations between the tetrahedral and a planar defect configuration as one possible dynamical process, or the appearance of polar order which leads to polar vortex and circulating band states \cite{SH_PRE_2015,NRV_PRF_2019}, or pure positional order as in active crystals \cite{praetorius2018active}, which can lead to collective rotation as seen in epithelia tissue confined on a sphere \cite{Tanner_PNAS_2012,Wang_PNAS_2013,HWV_EPL_2022,Tan2022.09.29.510101}. While in flat space activity can lead to continuous creation and annihiliation of defects \cite{Doostmohammadi_NC_2018}, in topological active matter such creation of defects is suppressed over a wide range of activity. 

\subsection{Phase field crystal and other modeling approaches}
In \cite{AWL_PRE_2011} a phase field crystal (PFC) model for liquid crystals was introduced, and due to various couplings between the rescaled density field, the local nematic order parameter, and the mean local direction of the orientations and their gradients, various stable phases of the equilibrium free-energy functional could be identified, including the smectic-A phase. Formulating this model on a sphere allows us to analyse these phases under the topological constraint and the influence of curvature. Besides the equilibrium configurations, the dynamics of these liquid crystalline phases is of importance, which could be addressed by the diffusive dynamics of the PFC model for liquid crystals. However, results mainly exist for systems with pure orientational or positional order, ranging from the classical Lifshitz-Slyozov-Wagner theory \cite{RadtkeVoorhees}, to large scale PFC simulations \cite{BGW11,Backofen2014} and simulations for nematic liquid crystals on surfaces \cite{Nitschkeetal_PRSA_2018,Nitschkeetal_PRSA_2020}. They are all characterized by annihilation of defects and convergence towards a (local) minimum with energetically prefered defects and the corresponding spatial arrangement. In addition, active PFC models \cite{menzel2013traveling,menzel2014active,Alaimoetal_NJP_2016,Alaimoetal_PRE_2018,Ophausetal_PRE_2018,Huangetal_PRL_2020,Ophausetal_PRE_2021,HLV_CP_2022} have been introduced. These models are related to the PFC model for liquid crystals \cite{AWL_PRE_2011} and have been formulated on a sphere in \cite{lowen2010phase,praetorius2018active} to study active crystals. 

\section{Results}

In the following we first introduce the surface active PFC model and then describe large-scale simulation results for active smectics on a sphere. We compare results of defect dynamics in multidomain smectic configurations, their coarsening and the emerging of stable configurations to the situation in flat space with periodic boundary conditions.

\subsection{Model}

In \cite{praetorius2018active} the active PFC model introduced in \cite{menzel2013traveling} has been formulated on a sphere $\Sp=R\,\Sphere$ with radius $R$, where $\Sphere$ is the two dimensional unit sphere embedded in $\mathbb{R}^3$. The model was used to study active crystals. Here, we consider a different parameter setting which is applicable for active smectics. In the model the position $\rb$ on the sphere $\Sp$ is parametrized by $\rb(\theta,\phi)=R \uu(\theta,\phi)$ with the orientational unit vector $\uu(\theta,\phi)=(\sin\theta\cos\phi,\sin\theta\sin\phi,\cos\theta)^{\mathrm{T}}$ and the spherical coordinates $\theta\in [0,\pi]$ and $\phi\in [0,2\pi)$. We consider a rescaled density field $\psi(\rb,t)$ and a polarization field $\pb(\rb,t)$, which is tangential to $\Sp$ at $\rb$, i.e., $\pb(\rb,t)=p_{\theta}(\rb,t)\partial_{\theta}\uu + p_{\phi}(\rb,t)\partial_{\phi}\uu \in\TS$ with scalar polarization fields $p_{\theta}(\rb,t)$ and $p_{\phi}(\rb,t)$ and the tangent space $\TS$ of the sphere $\Sp$ at point $\rb$. 

The free-energy functionals read
{\allowbreak\begin{align}%
\mathcal{F}_{\psi}[\psi] &= \int_{\Sp}\!\! \Big\{ \frac{1}{2}\psi\big[\TempR + (q_0 + \LaplaceS)^{2}\big]\psi + \frac{1}{4}\psi^{4} \Big\} \dif\surf \,, \label{eq:F_psi_S} \\
\mathcal{F}_{\pb}[\pb] &= \int_{\Sp}\!\! \Big( \frac{1}{2} C_{1} \norm{\pb}^{2} + \frac{1}{4}C_{2} \norm{\pb}^{4} \Big) \dif\surf \,, \label{eq:F_p_S}%
\end{align}}%
with $\epsilon < 0$, the characteristic wave number $q_0 = 1$ after rescaling, $C_1 > 0$ which leads to the suppression of any spontaneous ordering of orientational alignment, and $C_2$ modeling the higher order nonlinear term which is neglected in the following. The gradient and divergence operators are considered in spherical coordinates, i.e.,
{\allowbreak\begin{align}%
\GradS\psi &= \frac{1}{R}\Big[ (\partial_{\theta}\uu)\partial_{\theta}\psi
+ \frac{1}{\sin^{2}\theta}(\partial_{\phi}\uu)\partial_{\phi}\psi \Big],  
\\
\DivS\pb &= \frac{1}{R}\big( \cot\theta p_{\theta} + \partial_{\theta} p_{\theta} + \partial_{\phi} p_{\phi} \big),
\end{align}}%
and the surface Laplace-Beltrami operator is defined as $\LaplaceS=\DivS\,\GradS$. 

The dynamic equations describe the active-particle transport tangential to $\Sp$ and read
{\allowbreak\begin{align}%
\partial_{t} \psi &= \LaplaceS\frac{\delta\mathcal{F}_{\psi}}{\delta\psi} - v_{0}\,\DivS\pb \,, \label{eq:apfc_SI} \\
\partial_{t} \pb &= -(\LaplaceDR + D_{r})\frac{\delta\mathcal{F}_{\pb}}{\delta\pb} - v_{0}\,\GradS\psi \,, \label{eq:apfc_SII}
\end{align}}%
with $v_0$ the strength of particle self-propulsion and $D_r$ the rotational diffusion constant. The vector-Laplacian $\LaplaceDR$ is the surface Laplace-de\,Rham operator $\LaplaceDR=-\GradS\DivS-\rotS\RotS$, where
{\allowdisplaybreaks\begin{align}%
\rotS\psi &= \frac{1}{R\sin\theta}\big[-(\partial_{\theta}\uu)\partial_{\phi}\psi 
+ (\partial_{\phi}\uu)\partial_{\theta}\psi\big],
\\
\RotS\pb &= \frac{1}{R}\Big(2\cos\theta p_{\phi} \!-\! \frac{1}{\sin\theta}\partial_{\phi} p_{\theta} + \sin\theta \partial_{\theta} p_{\phi}\Big),
\end{align}}%
are the surface curl operators in spherical coordinates. The equations have been rescaled with a diffusive timescale and a length scale set via the pattern periodicity, and in this form constitute the minimal continuum field model which was used in \cite{praetorius2018active} to model active crystals on a sphere. When $v_{0}=0$, Eq. \eqref{eq:apfc_SI} reduces to the surface PFC model, which has been applied to describe optimal ordering and crystallization of passive particles on a sphere \cite{BackofenVoigtWitkowski_PRE_2010,Koehler_PRL_2016}. Equations \eqref{eq:apfc_SI} and \eqref{eq:apfc_SII} reduces to the active PFC model \cite{menzel2013traveling} in flat space in the limit $R\to\infty$. Within this limit, the model was used in \cite{HLV_CP_2022} to simulate defect dynamics in two-dimensional active smectics. This has been done by considering a parameter setting of $\epsilon = -0.98$ and the average density $\psi_0 = 0$ which favours a travelling stripe phase \cite{menzel2013traveling}. Such a stripe phase can also emerge on a sphere. Regarding the remaining model parameters we follow the aforementioned studies \cite{Menzel_PR_2015,praetorius2018active} and choose $D_r=0.5$, $C_1=0.2$, and $C_2=0$. For spherical geometry we consider a sphere with radius $R=100$, such that for passive case $50$ repetitions of a rotational symmetric smectic pattern perfectly fit to the surface. In the flat case a square periodic domain is simulated with size $L$, where $L$ is chosen to match domain area \wrt\ the sphere.

We use vector spherical harmonics, details in the methods section \ref{sec:methods}, to conduct multiple simulations ($15$ for flat geometry and $50$ for spherical geometry) of the coarsening process for each considered activity $\activity \in \{0.305,\, 0.35,\, 0.425,\, 0.5\}$. Each simulation starts from uniform $\psi(0) = \psi_0$ disturbed by random noise $\psi'$ with magnitude $|\psi'| < 0.01$ applied at each grid point.

In order to identify orientational defects we define a nematic order parameter almost everywhere $\xb \mbox{ : } \| \GradS \density(\xb) \| \neq 0$:
\begin{align}
    \qtensor(\xb) = \dirNematic(\xb) \dirNematic(\xb) - \frac{1}{2}\ProjSurf(\xb), \quad \dirNematic(\xb) = \frac{\GradS \density(\xb) \times \normalSurf(\xb)}{ \| \GradS \density(\xb) \|}, \label{eq:NematicOrderParameter}
\end{align}
with $\ProjSurf(\xb) = \mathbf{I} - \normalSurf(\xb)\normalSurf(\xb)$ the surface identity of the considered geometry and $\normalSurf$ the outward oriented surface normal. The nematic texture is then extended to the complete domain by smooth continuation. An associated distortion energy can be defined along the tensor components $I,J$ of $\qtensor$ \wrt\ the coordinates of embedding space $\mathbb{R}^3$
\begin{align}
    {\mathcal{F}}_{\qtensor}[\qtensor] = \int_{\Sp} \energyDensity_d(\qtensor) \dif\surf, \quad \energyDensity_d(\qtensor) = \frac{1}{2} \sum_{I,J=1}^3\| \GradS \qtensor_{I,J} \| ^2, \label{eq:LandauDeGennesDistortion}
\end{align}
which will be used to identify defects. For details we refer to the methods section \ref{sec:methods}.

\subsection{Pattern formation in active smectics}

\begin{figure}[ht!]
  \begin{center}
  \includegraphics[width=0.99\linewidth]{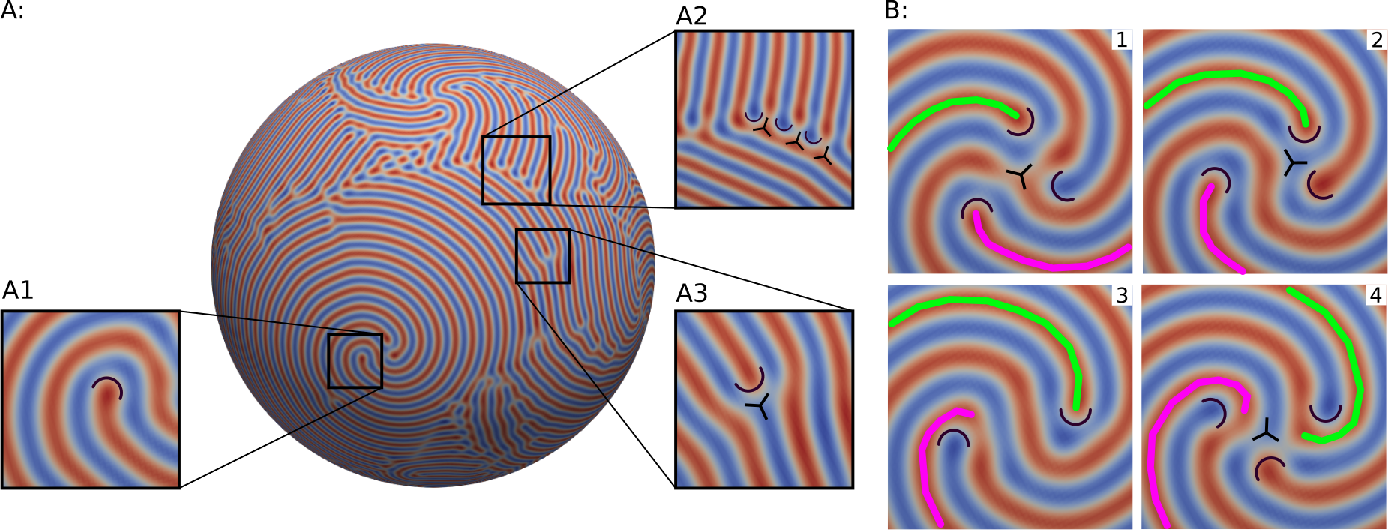}
  \caption{\textbf{Smectic pattern formation and dynamics in sphere domain.} [A] Snapshots of $\density$ exhibiting a typical set of phenomena with [A1] a single disclination in smectic line structure with topological charge of $+1/2$ (location denoted by black semi circle); [A2] a grain boundary, where the nematic texture exhibits an alternating pattern of $+1/2$ (black semi circles) and $-1/2$ (black tribars) defects; [A3] an isolated dislocation in smectic lines, for which the associated nematic texture consists of two disclination defects with opposite sign yielding an effective topological charge $0$ for the dislocation. [B] A rotating cluster formed by several disclinations, where panels 1--4 depict the temporal evolution of $\density$ with time steps of $\triangle t= 100$, showing an outward spiraling movement of lines (marked in green and pink) with rearranging disclinations and the associated nematic defects at center. The total topological charge of the cluster remains as $+1$ and its center does not move.}
  \label{fig:fig1}
  \end{center}
\end{figure}

\begin{figure}[ht!]
  \begin{center}
  \includegraphics[width=0.99\linewidth]{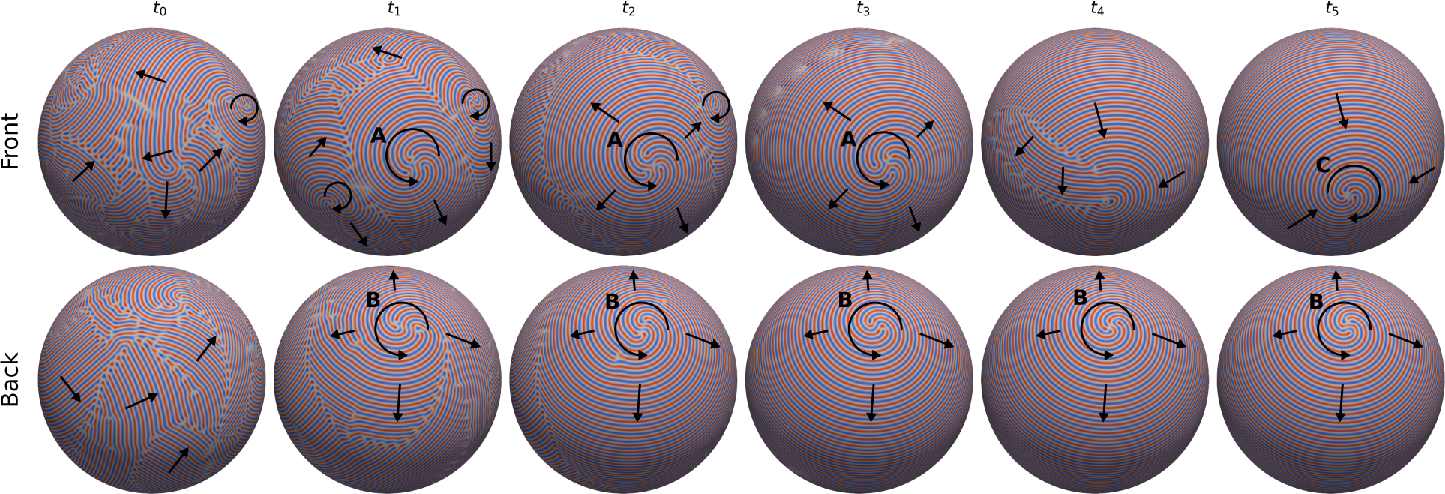}
  \caption{\textbf{Grain growth during the coarsening process for $\activity=0.5$ in sphere domain}, showing as a sequence of snapshots depicting $\density$ at various times marked in Simulation 4 of Fig. \ref{fig:qualiPhenomenaR100}, with front view (top panels) and back view (bottom) in terms of $180^{\circ}$ rotation \wrt\ vertical axis. Black arrows indicate the direction of grain movement and black circle arrows mark the rotating defect clusters. Black bold letters (\textbf{A}, \textbf{B}, \textbf{C}) indicate rotating clusters and correspond to the marks given in Fig. \ref{fig:fig3} [B]. Snapshots are taken at $t_0=500$, $t_1=900$, $t_2=1250$, $t_3=4000$, $t_4=9500$, and $t_5=20000$. An animation of the presented snapshots and intermediate grain configurations is provided in the SI.}
  \label{fig:grainDynamics}
  \end{center}
\end{figure}

In order to discuss the complex dynamics of smectic pattern formation and coarsening we introduce the following notions: 
\begin{itemize}
    \item \textit{Smectic lines} are lines of minimal (maximal) density $\psi$. In equilibrium they are quasi parallel and equidistant. In the active system these lines move consistently in a direction orthogonal to themselves; 
    \item \textit{Grains} are regions with quasi parallel and equidistant smectic lines. They are characterized by the orientation of the lines or the direction of their movement; 
    \item \textit{Grain boundaries} separate grains with different line orientation or movement direction from each other. If more than two grains are in contact the grain boundaries meet in junction points; \item \textit{Defects} are localized areas where the smectic lines are not parallel or not equidistant. For each defect a topological charge is associated; 
    \item \textit{Defect clusters} are compact aggregates of defects. In contrast to grain boundaries, which are typically chainlike aggregations of defects, these clusters lead to rotation; 
    \item \textit{Rotating grains} are a combination of defect clusters and their associated grains. Due to the rotation of the cluster, smectic lines form an Archimedes spiral with two rotation directions and two movement directions of the smectic lines. They move either inward or outward.
\end{itemize}
Fig. \ref{fig:fig1} and \ref{fig:grainDynamics} highlight most of these features. All phenomena introduced above can be observed in the video corresponding to Fig. \ref{fig:grainDynamics} (see SI). 

Fig. \ref{fig:grainDynamics} and the corresponding video in SI show the coarsening process for one simulation with $v_0 = 0.5$. Within Fig. \ref{fig:grainDynamics} at $t_0=500$ smectic lines have emerged and formed multiple grains. These grains move orthogonal to the smectic lines, with black arrows indicating the direction of movement. These phenomena are similar to the corresponding state in flat 2D space and are also reminiscent to the known coarsening processes in passive systems. At $t_1=900$, at the front side (top panel) grain boundaries curl up and form rotating grains which induce outward movement \wrt\ the central rotating defect cluster. On the back side (bottom panel) the direction of the grain movement changes without the curling up of grain boundaries. At $t_2=1250$ two counterclockwise rotating grains have formed with outward moving smectic lines, inducing a long grain boundary between these two incompatible grains. Note that the rotating defect clusters do not move as soon as they are formed. After a metastable state where grain boundaries wiggle, at $t_3=4000$ the rotating grain with its center at the back side expands and destroys the rotating cluster at the front side. Snapshot at $t_4=9500$ depicts an intermediate configuration where the front cluster is destroyed, yet the remaining grain moves in the clockwise direction and collapses. This process leads to the final single grain configuration at $t_5=20000$ consisting of a counterclockwise rotating cluster at the back side with outward moving smectic lines. On the opposite side, with maximum geodesic distance, a clockwise rotating cluster has formed with inward moving smectic lines. The corresponding transient of the distortion energy ${\mathcal{F}}_{\qtensor}$, together with the considered time instances, is shown in Fig. \ref{fig:qualiPhenomenaR100} (top panel of Simulation 4). 

\subsection{Distinct coarsening regimes}

\begin{figure}[ht!]
  \begin{center}
  \includegraphics[width=0.99\linewidth]{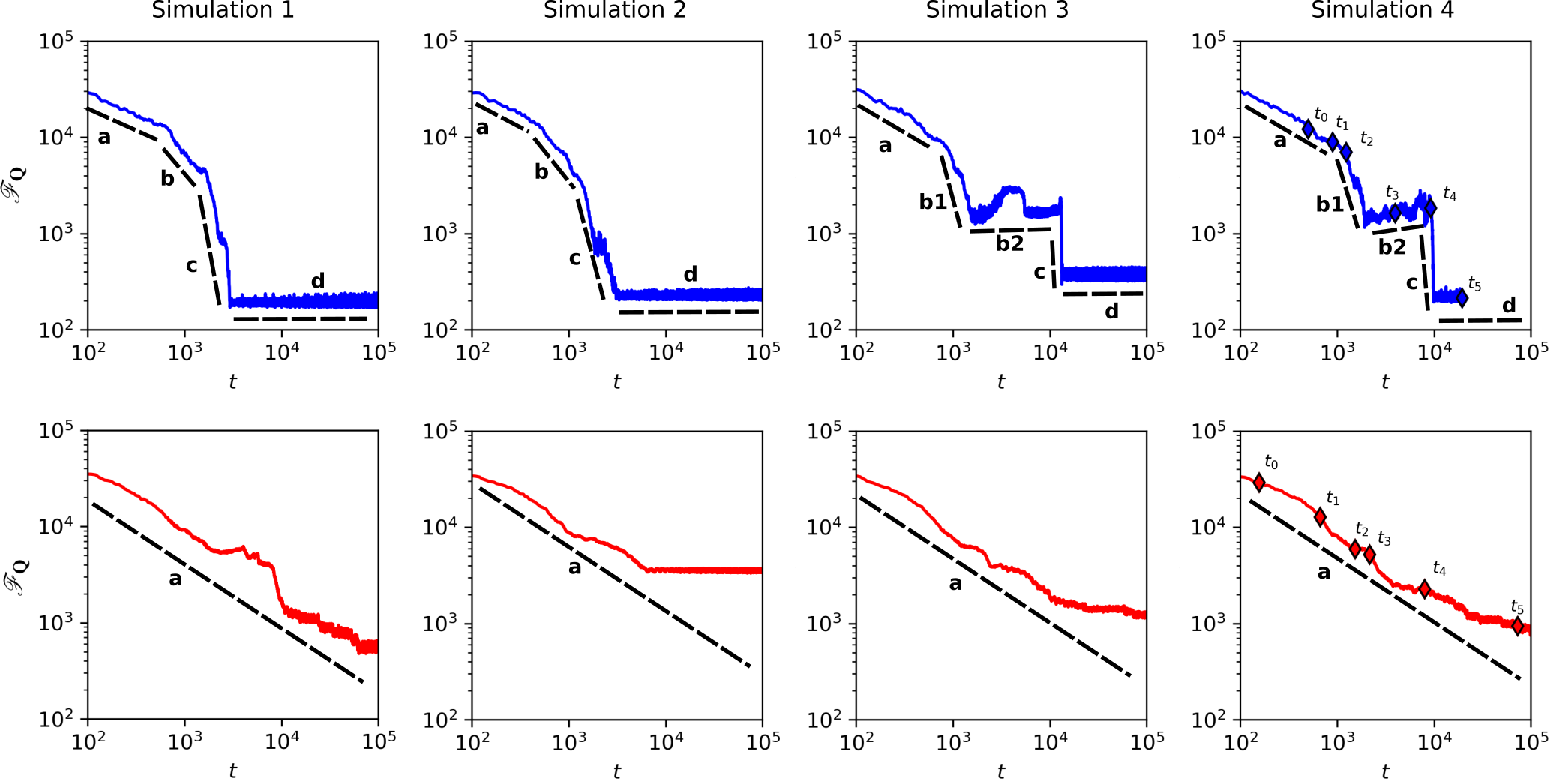}
  \caption{\textbf{Sample coarsening processes for $\activity=0.5$}: In sphere (top) and plane (bottom) domain, showing the time evolution of nematic distortion energy $\mathcal{F}_{\qtensor}(t)$ for four different simulations. Diamond markers in Simulation 4 correspond to the snapshots used in Fig. \ref{fig:grainDynamics} and Fig. \ref{fig:fig3}. Labels \textbf{a}, \textbf{b}, \textbf{c} and \textbf{d} mark distinct coarsening regimes. The scaling exponent for the plane domain is roughly $-1/2$ in all four simulations, as indicated by the dashed lines.}
  \label{fig:qualiPhenomenaR100}
  \end{center}
\end{figure}

\begin{figure}[ht!]
  \begin{center}
  \includegraphics[width=0.99\linewidth]{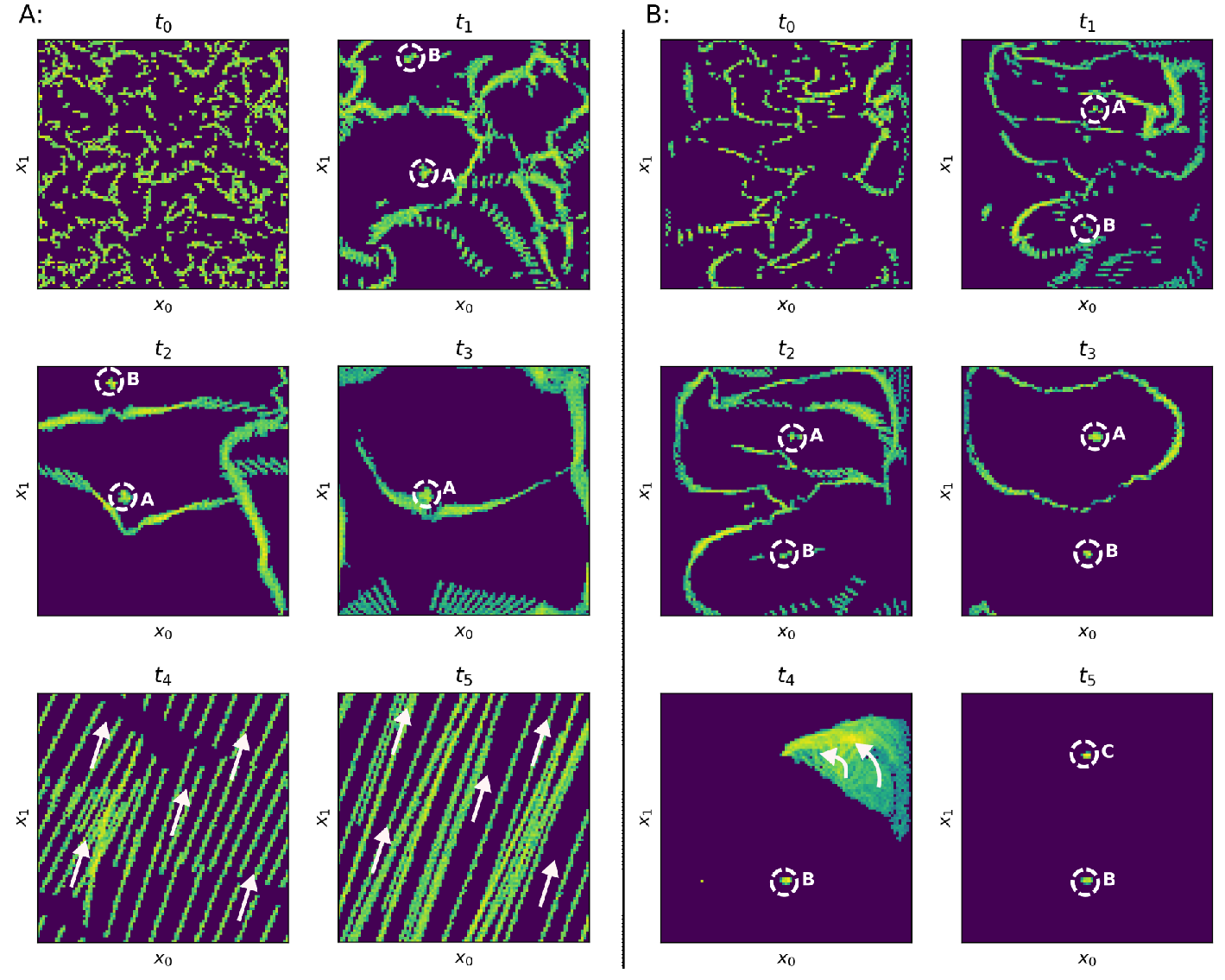}
  \caption{\textbf{Dynamics of defect distribution during the coarsening process:} Left panel [A] for planar and right panel [B] for sphere domain for $\activity=0.5$. The different time instances $t_0$, $\cdots$, $t_5$ correspond to the transients marked in Fig. \ref{fig:qualiPhenomenaR100} (Simulation 4) showing the nematic distortion energy $\mathcal{F}_{\qtensor}$. Dark blue patches indicate smectic lines that are free from nematic defects, while distinct grain boundaries are indicated by green to yellow lines. Movements of isolated defects are visible as short bright strokes. Rotating defect clusters and directions of defect motion are marked by white arrows. A video of the corresponding time evolution is provided in the SI.}
  \label{fig:fig3}
  \end{center}
\end{figure}

The transients of the distortion energy ${\mathcal{F}}_{\qtensor}$ in the spherical case, as shown in Fig. \ref{fig:qualiPhenomenaR100} (top panels), indicate different coarsening regimes. While different simulation runs differ quantitatively, they share common qualitative features. 
We can identify four distinct regimes which are marked in each plot by labels \textbf{a}, \textbf{b}, \textbf{c} and \textbf{d}. 
Regime \textbf{a} refers to initial coarsening, as characterized by small grains which merge or dissolve. Defect clusters are rarely observed in this regime. The scaling exponent in this regime is roughly $-1/2$ in all simulations. In regime \textbf{b}, a set of large rotating clusters is formed, while small grains without rotating clusters are rarely present. The spatial arrangement of the clusters and their rotation direction seem random but impact the subsequent evolution. This regime is characterized by a coarsening of rotating clusters. As in Ostwald ripening, large rotating clusters grow on the expense of small rotating clusters. This leads to a faster coarsening compared to regime \textbf{a}. The associated rotating grains are more stable than grains without rotating clusters. A critical state is reached when a grain boundary reaches a rotating cluster and dissolves it. The remaining grain is then typically dissolved into the neighbouring rotating grain. Some intermediate configurations during the transition from regime a to regime \textbf{b} are more stable or long-lasting than others. In the configuration shown in Fig. \ref{fig:grainDynamics}, after a similar coarsening process indicated by \textbf{b1} in Fig. \ref{fig:qualiPhenomenaR100}, two almost equally sized rotating grains cover the domain, although their orientations are different, with a long grain boundary separating them. Due to their similar size the grain boundary migrates slowly, yielding a long time period with quasi stable configuration, as indicated by \textbf{b2} in Fig. \ref{fig:qualiPhenomenaR100}. In regime \textbf{c}, after the motion of grain boundary of the largest rotating grain has dissolved any other rotating clusters, only a single grain and the corresponding rotating cluster remain. The grain boundary then quickly collapses, and an inward-directed rotating defect cluster appears on the opposite pole position of the dominant rotating defect cluster. The collapse of the grain boundary happens very fast in a zipping mode, leading to a large coarsening exponent (see Section \ref{sec:fast_coarsening}). The final regime \textbf{d} is featured by a stable configuration of two rotating defect clusters (one inward, one outward) positioned on opposing sides/poles of the sphere with maximized distance. The resulting rotation axis is determined by the dominant rotating defect cluster emerging during the transition from regime a to regime \textbf{b}. Also, the number of defects in these final-state rotating defect clusters can vary in different simulations and is determined by the formation of the dominant rotating defect cluster. 

In order to unveil the qualitative difference of the plane vs. sphere domain, we conduct the same type of simulations on a planar 2D domain. The corresponding distortion energy is shown in Fig. \ref{fig:qualiPhenomenaR100} (bottom panels). They do not lead to distinct coarsening regimes and essentially can be described by a single regime a, with scaling exponent roughly $-1/2$ in all simulations (as indicated in the bottom panels), which terminates into the final configuration \textbf{d}.

To further understand the differences we visualize the dynamics of defect distributions in the temporally binned data around different time instances for one sample simulation, as shown in Fig. \ref{fig:fig3}, and compare the configurations on the sphere and the plane domains. For the sphere the considered time instances correspond to those of Fig. \ref{fig:grainDynamics}. These defect distributions are obtained by collecting defect positions around the selected time instances $t_0, \ldots, t_5$, each of which corresponds to one of the $100$ temporal bins (logarithmic sizes) set up between $t=10^2$ and $t=10^5$, and evaluating a spatial distribution $D_i(x_0,x_1)$ ($100\times 100$ spatial bins) of the positions. The resulting two-dimensional probability distribution of defect positions is plotted as log-scale colormap (dark blue to yellow) with range limited to $[10^{-3},1]$. Dark blue patches indicate smectic lines that are free from nematic defects, while distinct grain boundaries are indicated by green to yellow lines. Movement of isolated defects (disclinations or dislocations) is visible as short bright strokes. Rotating clusters are marked and indicated by dashed circles. On the sphere local coordinates $(x_0, x_1)$ are derived from spherical coordinates ($x_0$ corresponds to polar angle $\theta$, $x_1$ to azimuthal angle $\phi$) such that strong distortions can appear near left and right boundaries of snapshots. 

Initially the situation is similar on the sphere and the plane domain. At the early time $t_0$ an initial establishment of smectic lines in small patches is visible for both domains. These patches merge and rearrange to larger grains at time $t_1$. For the sphere domain, around time $t_1$ two rotating clusters are formed, marked as \textbf{A} and \textbf{B}. This occurs by the curling of grain boundaries. At a later time $t_2$ they evolve to two similarly sized grains, yielding a metastable state where grain boundaries hardly move. Eventually the grain boundaries move towards cluster \textbf{A} at $t_3$. In the $t_4$ snapshot cluster \textbf{A} has been dissolved and the grain boundary collapses with another rotating cluster. At $t_5$ the final minimal configuration (ground state dictated by topology) of an outward rotating cluster \textbf{B} and an inward rotating cluster \textbf{C} is formed. This evolution process corresponds to the snapshots given in Fig. \ref{fig:grainDynamics}. Especially the collapse of the grain boundary at $t_4$ is characteristic of the fast decay of $\mathcal{F}_{\qtensor}$ shown in Fig. \ref{fig:qualiPhenomenaR100}. 

The situation differs on the plane domain. At $t_1$ there are two grains each containing a rotating cluster, labeled \textbf{A} and \textbf{B}. In transition from $t_1$ to $t_2$ grains expand and merge, and only the grains containing rotating clusters \textbf{A} and \textbf{B} remain. At $t_3$ the moving grain boundary reaches a rotating cluster and dissolves it (as happened for cluster \textbf{B} in transition from $t_2$ to $t_3$). With the vanishing of the last rotating cluster \textbf{A} the grain boundaries dissolve into isolated defects and the collective motion of smectic lines starts. In the $t_4$ snapshot parallel bright strokes indicate collective movement of isolated defects along the direction indicated by arrows. They rearrange and move collectively at the final configuration shown at $t_5$. 
Identifying these details for the other simulation runs of Fig. \ref{fig:qualiPhenomenaR100} leads to the same qualitative features. We never observe a collapsing grain boundary in a planar 2D geometry as on the sphere domain at $t_4$. We therefore associate this phenomenon of grain collapse with the topological and geometrical constraints of the sphere.

\subsection{Fast coarsening regime}
\label{sec:fast_coarsening}

\begin{figure}[ht!]
  \begin{center}
  \includegraphics[width=0.99\linewidth]{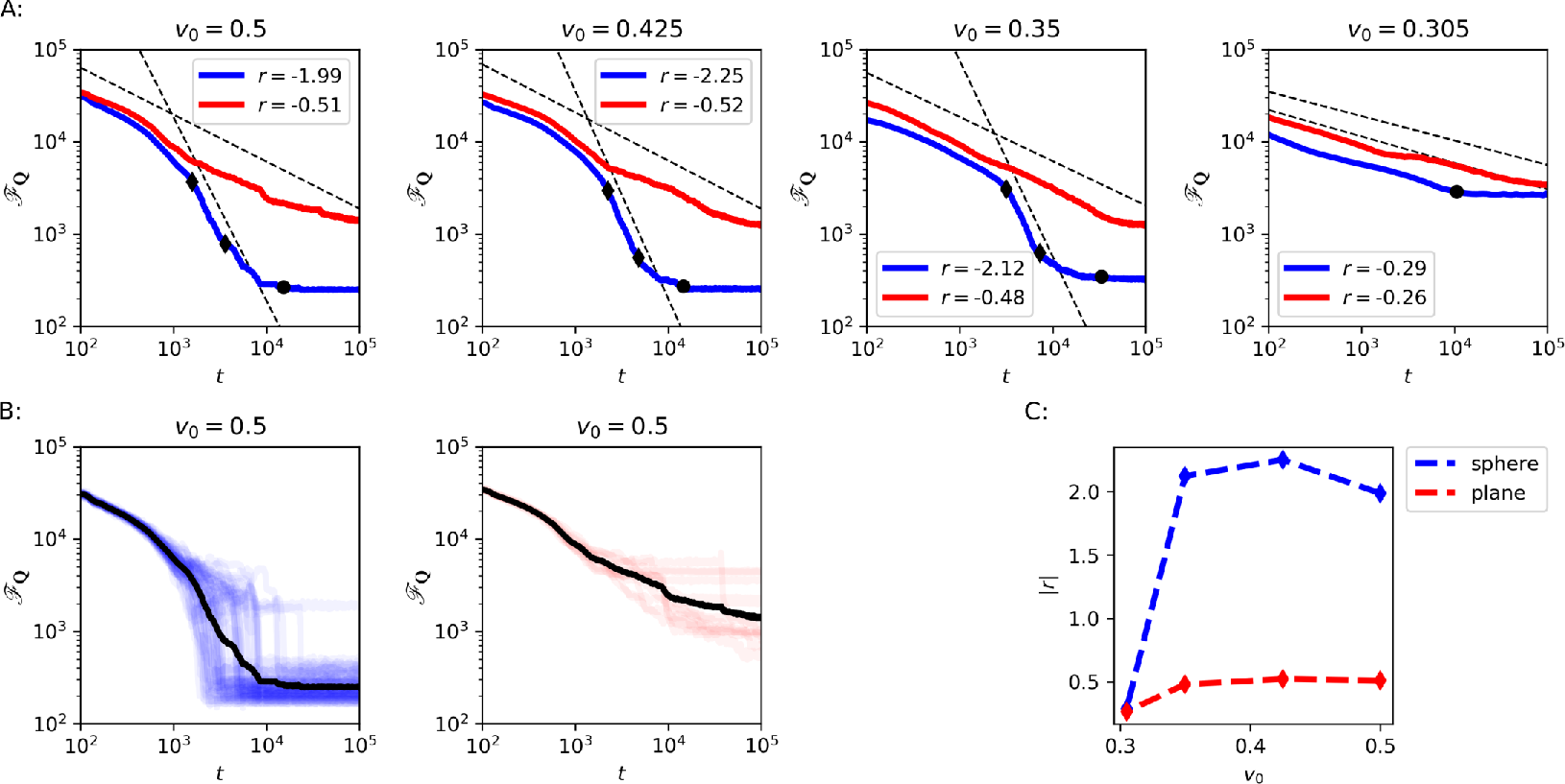}
  \caption{\textbf{Average behavior of coarsening for sphere and plane domains.} [A] The nematic distortion energy $\mathcal{F}_{\qtensor}$ averaged over ensemble of simulation runs on sphere (blue lines, 50 runs for each activity) and plane (red lines, 10 runs for each activity). Black dashed lines indicate the fitted power-law decay. Black diamonds and dots indicate the fitting interval for the fast coarsening regime and the transition to the steady state. [B] Time evolution of $\mathcal{F}_{\qtensor}$ at $\activity=0.5$, for individual runs (light color lines) and the corresponding average (solid black line). Left and right panels correspond to sphere and plane geometry, respectively. [C] Summary of values of scaling exponent $r$ obtained from the fitting in [A] as a function of activity $v_0$, for sphere (blue) and plane (red) geometries.}
  \label{fig:fig2}
  \end{center}
\end{figure}

We next examine how the fast coarsening regime identified above is influenced by the activity and how it influences statistical coarsening laws. We therefore conduct more simulations at different values of $v_0$, and again compare our results with the corresponding simulations in flat space. Fig. \ref{fig:fig2} [A] shows the results of averaged distortion energy ${\mathcal{F}}_\qtensor$ over time, which can be related to the statistics on the time decay of defect density. 

For the planar geometry there is no clear distinction between different scaling regimes, as shown in both Fig. \ref{fig:qualiPhenomenaR100} (bottom panels) and Fig. \ref{fig:fig2} [A]. This is similar to passive smectics in 2D space. At small activity $v_0 = 0.305$ we obtain the value of coarsening exponent of $0.26$, close to the scaling of $t^{-1/3}$ for dislocations in 2D passive smectics \cite{BV_PRE_2002,AR_NJP_2008}. At large enough activity the scaling is roughly $t^{-1/2}$. 

For the sphere the initial coarsening regime, i.e., regime \textbf{a} in Fig. \ref{fig:qualiPhenomenaR100}, shows a similar scaling behavior as the flat geometry. At low activity this regime spans over most of the time range until a steady state with low defect density is reached, analogous to the behavior of the planar geometry (see the blue and red lines in Fig. \ref{fig:fig2} [A] at $v_0 = 0.305$, where the black dot indicates the transition towards the steady state configuration). The topological constraint and the curvature of the sphere therefore do not affect the grain coarsening. However, at high enough activity the dynamical process of coarsening changes qualitatively, for which a transition to a faster coarsening regime becomes evident. The transition into this regime happens earlier with increased activity, as seen in Fig. \ref{fig:fig2} [A] ($v_0 = 0.5$, $0.425$, and $0.35$; blue lines). Much faster coarsening processes, with much larger scaling exponents, are observed, with a scaling behavior of roughly $t^{-2}$ (see Fig. \ref{fig:fig2} [A,C]). These scaling exponents are calculated for the time range combining regimes \textbf{b} and \textbf{c} of  \ref{fig:qualiPhenomenaR100}. The transition between these regimes strongly depends on the specific configurations, as seen in  \ref{fig:fig2} [B] showing results from all the individual simulation runs on the sphere and the flat domains for $v_0 = 0.5$. As the activity increases, the transition into the fast coarsening regime occurs at earlier time, and the final configuration is reached earlier as well (see the black diamonds in Fig. \ref{fig:fig2} [A] at $v_0 = 0.5$, $0.425$, and $0.35$, which mark the start and end times for the fast coarsening regime, and the black dots which mark the transition to the steady state).

\subsection{Testing defect avalanche phenomenon}

One could ask if the grain boundary collapse within this fast coarsening regime shows any features of avalanche phenomenon. To explore this we follow the characterization of avalanche event given in \cite{chan2010plasticity}, through the number of defects $\numDefect(t)$ and the defect velocity $\defectVelo_d$, and a sudden change of both quantities. Within an avalanche the defect velocity is significantly larger than during the whole coarsening process and the number of defects reduces faster. The later has already been identified by the time evolution of $\mathcal{F}_{\qtensor}$ in Fig. \ref{fig:fig2}, which relates to the defect density and thus also to the number of defects. Here we examine the velocity of the defects in this fast coarsening regime, with the comparison between the sphere and plane domains for $v_0 = 0.5$.

\begin{figure}[ht!]
  \begin{center}
  \includegraphics[width=0.86\linewidth]{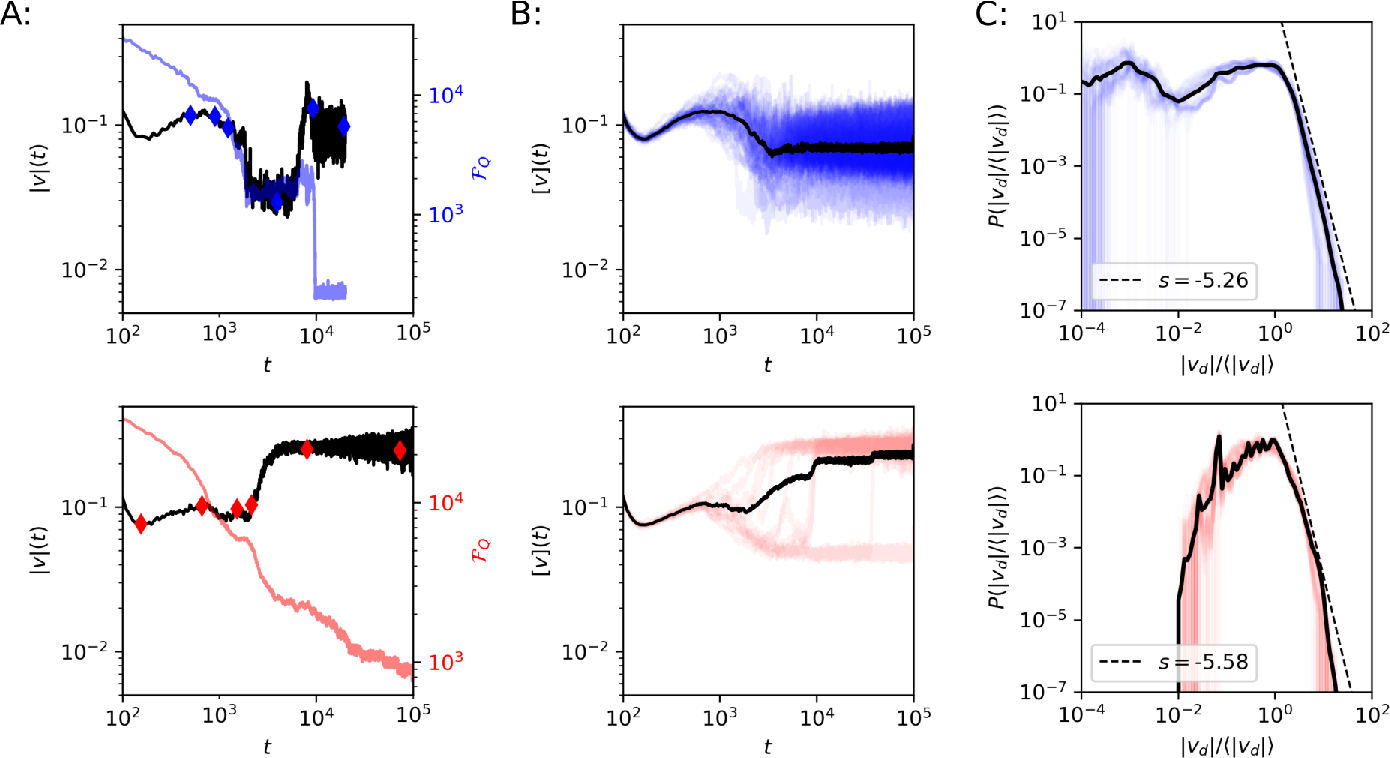}
  \caption{\textbf{Statistics of defect velocities:} for sphere (top) and plane (bottom) domains at a large activity $\activity=0.5$. [A] Evolution of average defect velocity magnitude determined at each simulation timestep $|v|(t) = \frac{1}{\numDefect(t)}\,\sum_{n=1}^{N(t)} \|v_d(t,n)\|$. This data corresponds to Simulation 4 in Fig. \ref{fig:qualiPhenomenaR100} and is shown together with the distortion energy $\mathcal{F}_{\qtensor}$ (blue or red curve, where diamonds mark the times of snapshots shown in Fig. \ref{fig:grainDynamics}). [B] Ensemble average of defect velocity magnitude $[v](t) = \frac{1}{M} \sum_{m=1}^M \frac{1}{\numDefect_m(t)} \sum_{n=1}^{\numDefect_m(t)} \| v_d(t,n,m) \| $ across all simulations ($M=50$ for sphere geometry, $M=10$ for flat geometry), together with the individual plots of $|v|(t)$ for all $M$ runs (blue and red lines). [C] Probability distribution for defect velocity magnitude for combined data set across all the simulations and the complete time domain $t\in[10^2,10^5]$ (black line), and for temporal binned data (blues and red lines) where $[10^2,10^5]$ is separated into $50$ slices of logarithmic scaled boxes (as in \cite{angheluta2012anisotropic}). The power law fitting, with the obtained exponent $s$, for high velocities $|v_d|/\langle |v_d| \rangle \geq 1$ is marked by black dashed line.}
  \label{fig:fig4}
  \end{center}
\end{figure}

Fig. \ref{fig:fig4} shows the evolution of the average defect velocity magnitude for Simulation 4 of Fig. \ref{fig:qualiPhenomenaR100}, its ensemble average over all simulations, and the probability distribution of this data over the complete time domain and of temporal binned data. All results are shown for the sphere and the plane domain. For high velocities $|v_d|/\langle |v_d| \rangle \geq 1$ a power law scaling with an exponent of roughly $-5.5$ can be identified for the probability distribution for both the sphere and plane domains; see Fig. \ref{fig:fig4} [C]. This strong decay cannot be exclusively attributed to the fast coarsening regime. First, the distributions are similar for the sphere and plane domains at high velocities while the fast coarsening regime is not present for the planar domain. Second, the probability distribution of the considered time bins, i.e., the blue lines in Fig. \ref{fig:fig4} [C], shows essentially the same behaviour as the overall distribution, which indicates that the high velocities are not limited to any specific time span within the coarsening process. This is confirmed by the time evolution of the averaged defect velocity magnitude in Fig. \ref{fig:fig4} [A,B], which does not show a significantly larger velocity around the time instance $t_4$ (fast coarsening regime). We only obtain a plateau of lower velocities before the fast coarsening regime. Thus, these results are not sufficient to identify the grain boundary collapse observed in the fast coarsening regime as an avalanche phenomenon.

\section{Discussion and Conclusions}

The complex interplay of orientational and positional defects under topological constraints, geometric effects, and active driving has been investigated for active smectics on a sphere using an active surface PFC model. Above a critical activity threshold we have identified a stable state of rotating spirals which emerge from two rotating defect clusters. While the number of defects in these clusters varies, they are always localized and positioned to maximize the distance between each other. Such spiral states are reminiscent to spirals in reaction-diffusions on a sphere \cite{MS_N_1989,MG_JPC_1996,GA_PRE_1997,DGK_PRE_2004,AG_JCAM_2005,CHL_SIAMR_2008,SM_SIAMJADS_2011}.

The coarsening process that the system undergoes before it reaches this stable state significantly differs from self-similar coarsening in passive smectics or active smectics in the planar geometry. While several features of grain boundary motion at early coarsening are similar to the passive case, differences associated with activity are already present in this early regime. These differences are rotating defect clusters, which lead to an enhanced coarsening with defect density scaling of roughly $t^{-1/2}$ instead of roughly $t^{-1/3}$ in the passive case. This increase in the coarsening rate can be associated with activity and is independent of the considered geometry of sphere or planar domain. A key characteristic for active smectics on a sphere is the appearance of a fast coarsening regime. It is present at late coarsening, just before the steady state is reached. This regime is characterized by a scaling of roughly $t^{-2}$ for the time evolution of defect density and is only present on the sphere domain. It is associated with the collapse of grain boundary of a rotating grain. This collapse forms a new defect cluster which is located on a pole position opposite to the already present rotating defect cluster and rotates in the opposite direction, forming the rotating spiral state. This process of grain boundary collapse and fast coarsening is found to be different from the avalanche phenomenon. The velocity of the defects within this regime is not significantly larger than that during the whole coarsening process. Our simulations also show that this fast coarsening regime follows a time period of frustration, associated with a plateau or even an increase in the transient of the distortion energy $\mathcal{F}_{\qtensor}$ and significantly lower defect velocities. The extension of this plateau strongly depends on the local configuration.

This investigation on a sphere can only be the first step to understand the complex interplay in the coarsening process of active surface smectics. For surfaces with the same topology but varying curvature we speculate that a similar steady state configuration will be reached. However, varying surface curvature potentially localizes the rotating defect clusters at extrema of surface curvature, e.g., in prolate configurations, or even leads to more complicated metastable states for more complex geometries. We further speculate that a similar collapse of defects will be also observed on more general geometries just before the steady state configuration is reached. However, the observed plateau in the transient of the distortion energy before this event will probably strongly depend on local geometric features. All these speculations ask for further numerical investigation and experimental realization. Numerical methods which deal with non-spherical geometries and surface PFC or surface liquid crystals (polar and nematic) have been developed \cite{Aland1,Aland2,nestler2018orientational,Nitschkeetal_PRSA_2018} and need to be extended to the active case. Also, the evolution of the surface in response to the surface liquid crystals on it is an extension that can be studied numerically \cite{Nitschkeetal_PRSA_2020}. Extending this to surface smectics will allow us to consider surface instabilities, such as the Helfrich-Hurault elastic instability, which if considered for smectic shells can lead to surface undulations and buckling \cite{blanc2023helfrich}. Combined with activity this might be the path to model and simulate some biological applications (e.g., thosed mentioned in \cite{JPT_PRE_2022}) as active surface smectics.

\section{Methods} \label{sec:methods}

\subsection{Numerical solution by (vector) spherical harmonics}
We expand $\psi$ and $\pb$ in (vector) spherical harmonics so that Eqs. \eqref{eq:apfc_SI} and \eqref{eq:apfc_SII} reduce to a set of ordinary differential equations for the time-dependent expansion coefficients of $\psi$ and $\pb$. 
Let $\mathcal{I}_{\Nsh}=\{(l,m)\,:\,0\leq l\leq \Nsh,\,|m|\leq l\}$ be an index set of the spherical harmonics $Y^{m}_{l}:\Sp\to\C$ up to order $\Nsh$. The scalar field $\psi$ can be expanded as
\begin{equation}
\psi(\rb,t) = \!\!\! \sum_{(l,m)\in\,\mathcal{I}_{\infty}} \!\!\! \hat{\psi}_{lm}(t) Y_l^m(\rb),
\end{equation}
with the expansion coefficients $\hat{\psi}_{lm}(t)$. For the vector field $\pb:\Sp\to\mathrm{T}\Sp$ we consider the decomposition $\pb(\rb,t) = \GradS p_{1}(\rb,t) + \rotS p_2(\rb,t)$, and construct a tangent vector field basis from the gradient $\GradS$ and curl $\rotS=\uu\times\GradS$ of the spherical harmonics basis functions. We consider the vector spherical harmonics 
$\mathbf{y}_{lm}^{(1)}(\rb) = R\, \GradS Y_l^m(\rb)$ and $\mathbf{y}_{lm}^{(2)}(\rb) = -\frac{\rb}{\norm{\rb}}\times \mathbf{y}_{lm}^{(1)}(\rb)$, and represent $\pb$ by
\begin{equation}
\pb(\rb,t)=\sum_{i=1}^2\sum_{(l,m)\in\,\mathcal{I}_{\infty}} \!\!\! \pbc_{lm}^{(i)}(t) \mathbf{y}_{lm}^{(i)}(\rb),
\end{equation}
with expansion coefficients $\pbc_{lm}^{(i)}(t)$. The resulting Galerkin scheme \cite{Hesthaven2007} reads
\begin{align}%
\begin{split}%
\partial_{t} \hat{\psi}_{lm}(t) + \frac{l(l+1)}{R^2}\Big\{\TempR+\Big[1-\frac{l(l+1)}{R^2}\Big]^2\Big\}\hat{\psi}_{lm}(t) + \frac{l(l+1)}{R^2}\hat{\nu}_{lm}(t) - v_{0} \frac{l(l+1)}{R}\pbc_{lm}^{(1)}(t) &= 0 \,,
\label{eq:discrete_pfc}%
\end{split}\\
\begin{split}%
\partial_{t} \pbc_{lm}^{(i)}(t) + \Big[\frac{l(l+1)}{R^2} + D_r\Big]\big(C_{1}\pbc_{lm}^{(i)}(t) + C_{2}\qbc_{lm}^{(i)}(t)\big) + v_{0} \frac{\delta_{i1}}{R}\hat{\psi}_{lm}(t) &= 0, 
\label{eq:discrete_polar}%
\end{split}
\end{align}
with $i\in\{1,2\}$, $(l,m)\in\mathcal{I}_{\Nsh}$, and $t\in[t_{0},t_{\mathrm{end}}]$. Here $\hat{\nu}_{lm}(t)$ are the expansion coefficients of $\nu = \psi^3$, $\qbc_{lm}^{(i)}(t)$ are the expansion coefficients of $\qb = \norm{\pb}^2\pb$, 
and $t_{\mathrm{end}}$ represents the simulated time range starting at $t_{0}=0$. Quadrature on the sphere $\Sp$ is realized by evaluating $\psi$ and $\pb$ in Gaussian points $\{(\theta_{i}, \phi_{j}) \,:\, 1\leq i \leq\Nth, 1\leq j \leq\Nphi\}$, where $\Nth$ and $\Nphi$ are the numbers of grid points along the polar and azimuthal coordinates, respectively. An appropriate quadrature rule \cite{Schaeffer2013} is applied.
We use a second-order accurate scheme similar to that described in \cite{Backofen2011} for the time-discretization of Eqs.~\eqref{eq:discrete_pfc} and \eqref{eq:discrete_polar}. The implementation of the vector spherical harmonics is based on the toolbox SHTns \cite{Schaeffer2013} and identical to the one used in \cite{praetorius2018active}.

For numerical solution procedure we discretize the spherical domain with $N_{\theta} = 512$ and $N_{\phi}=1024$, and used spherical harmonics expansion of order $n=500$ and a time step $\tau=0.05$ for time-discretization. The simulations are performed on the time domain $[0, 10^5]$.

\subsection{Identification and tracking of disclinations and dislocations} To detect disclinations and dislocations in the smectic pattern we use the nematic order parameter of Eq. \eqref{eq:NematicOrderParameter} and its distortion energy defined in Eq. \eqref{eq:LandauDeGennesDistortion}. For these quantities several techniques have been proposed (see, \eg, \cite{wenzel2021defects}) to identify dislocations and disclinations. Here we use the approach to identify defects as localized maxima of distortion energy in Eq.~\eqref{eq:LandauDeGennesDistortion}. It is noted that in this context the distinction between dislocation and disclination is dropped. 
To exclude faulty detections due to numeric fluctuations in $\energyDensity_d(\qtensor)$ we use a lower threshold $\energyDensity_d(\qtensor) > 1$. Furthermore we filter the resulting positions by a distance threshold of $\| \xb_i - \xb_j \|_{\mathbb{R}^3} > \pi/2$ to reduce possible sets of maxima in $\energyDensity_d(\qtensor)$ close to a disclination to a single position.

At the next step we aim to identify larger coherent structures, like rotating defect clusters and grain boundaries. We evaluate a hierarchical clustering \cite{mullner2013fastcluster} of the positions along Euclidean distance with the requirement of single connection, using the implementation provided by \texttt{scipy.cluster} \cite{2020SciPy-NMeth}. The coherent structures are given by a slice (of the defect position dendrogram) with threshold $3/2 \pi$. As the final step we connect the instantaneous defect positions and clusters by using the \texttt{trackpy} \cite{allan2019soft} particle tracking library. From these trajectories we obtain defect velocity magnitude by central difference $\defectVelo_{d}(t^i) =[ \|\xb^{i} -\xb^{i-1} \|/(t^i -t^{i-1}) + \| \xb^{i+1}- \xb^{i}\|/(t^{i+1} - t^i) ]/2 $.

\subsection{Fast coarsening regime and steady state regime tracking} 
To quantify and track the fast coarsening regime we consider the distortion energy transient $(t, \mathcal{F}_{\qtensor}(t))$ in the $\log-\log$ space $(\underline{t} = \log_{10} t$, $\underline{\mathcal{F}}_{\qtensor} = \log_{10}\mathcal{F}_{\qtensor} )$. There we separate the $\log-\log$ transient in equidistant slices $[\underline{t}_i, \underline{t}_i+ \delta t], \, i \in \{0, \hdots 250 \}$, evaluate the relative energy decrease in each slice $d_i = [ \underline{\mathcal{F}}_{\qtensor}(\underline{t}_i+ \delta t) - \underline{\mathcal{F}}_{\qtensor}(\underline{t}_i)] / \max_k [ \underline{\mathcal{F}}_{\qtensor}(\underline{t}_k+ \delta t) - \underline{\mathcal{F}}_{\qtensor}(\underline{t}_k) ]$ and a threshold by $d_i > 0.75$ to obtain time slices containing fast coarsening. In a second step neighbouring slices with fast coarsening regime are combined.

Applying this method to the transients of averaged energies, as described in Fig. \ref{fig:fig2}, we obtain a single time domain with fast coarsening regime. Note that when applying such a method to an energy transient of a single simulation run we have to reduce noise in the transient, \eg, by convolution with a Gaussian function along temporal axis in the $\log-\log$ space, and then yield possibly more than one time domains with fast coarsening regime.

To identify a steady state regime we determine the average nematic distortion energy $\langle \mathcal{F}_{\qtensor} \rangle$ in the last $5\%$ of simulation time domain $[0,T]$. If the energy at $t=0.95T$, the begin of averaging domain, is less than a value $5\%$ higher than the averaged energy, i.e., if $\mathcal{F}_{\qtensor}(0.95 T) < 1.05 \langle \mathcal{F}_{\qtensor} \rangle$, we classify the simulation as reaching the steady state. If a steady state is reached, we find the steady state domain $[T_s, T]$ by traversing the simulation time steps backward, starting from $t = 0.95T$, as long as the criteria $\mathcal{F}_{\qtensor}(t) < 1.05 \langle \mathcal{F}_{\qtensor} \rangle$ is satisfied.

\subsection{Decay rate fitting} 
To estimate the power-law decay rates as done in Figures \ref{fig:fig2} and \ref{fig:fig4}, we transform the considered data into the $\log-\log$ space and perform linear regression on the relevant subset of the data to obtain the decay rate $r$. For Fig. \ref{fig:fig2} we use the full data set in cases where no steady state was tracked, while in cases with steady state we consider data sets either restricted to $[0,T_s]$ (if no fast coarsening regime was tracked) or restricted to the regime of fast coarsening. In Fig. \ref{fig:fig4} we restrict the data set to velocities faster than the average, i.e., $|v_d|/\langle |v_d| \rangle \geq 1$. 

\subsection{Statistics of defect velocities}
As we are concerned with evolution in a time span across several orders of magnitude, we typically describe the phenomena in $\log-\log$ space. To reduce noise in the simulation results we apply the temporal binning with $\log$ sized bins as provided by \texttt{numpy.logspace}\cite{harris2020array} such that the considered bins are equally sized in $\log$ space.

To characterize the defect velocity we apply several notions reflecting the possible perspectives on the data. As the first notion we consider $$|v|(t) = \frac{1}{\numDefect(t)} \sum_{n=1}^{\numDefect(t)} \|v_d(t; n,I)\|, $$ the average defect velocity magnitude across a defect configuration (consisting of $\numDefect(t)$ defects) at a certain time $t$ and in a simulation $I$. To obtain a characteristic defect velocity magnitude independent of instantaneous defect configuration and specific realization of evolution, which highly depend on the random initial value, we introduce an ensemble average across $M$ simulations to obtain the characteristic average defect velocity magnitude $$[v](t) = \frac{1}{M} \sum_{m=1}^M \frac{1}{\numDefect_m(t)} \sum_{n=1}^{\numDefect_m(t)} \| v_d(t,n,m) \|.$$ Furthermore, we consider the probability distribution of defect velocities to obtain a notion compatible to \cite{angheluta2012anisotropic}. For this purpose we consider the magnitude of defect velocity as a random variable. We calculate the associated expected value $\langle |v_d| \rangle$ and probability distribution numerically from the combined data set of all defect configurations across all simulations, $\|v_d(t,n,m)\|, \forall n=1\hdots N,\,m=1\hdots M,\, t\in[10^2,10^5]$, as shown in Fig. \ref{fig:fig4} [C] (black curves). To test whether the observed different coarsening regimes have an impact on this probability distribution we repeat the calculation by restricting the defect velocity data to different temporal bins with the use of 50 slices of logarithmic scaled boxes. 

\section*{References}
\bibliographystyle{iopart-num-mod.bst}
\bibliography{references}

\section*{Acknowledgements}

We are thankful for the support of this work by the Deutsche Forschungsgemeinschaft (DFG) through the grants LO 418/20-2 (for H.L.) and VO 899/19-2 (for A.V.), PR 1705/1-1 (for S.P.) within the DFG Research Unit 3013, and the National Science Foundation (NSF) under Grant No. DMR-2006446 (for Z.-F.H.). We further acknowledge computing resources provided at FZ J\"ulich within grant pfamdis and at ZIH/TU Dresden within grant WIR. 

\section*{Author contributions statement}

Z.-F.H., H.L. and A.V. developed the model and conceived the computational experiments. S.P. implemented the simulation algorithm for the spherical geometry, and Z.-F.H implemented the simulation algorithm for the planar case. M.N. implemented the postprocessing algorithms and conducted the computational experiments. M.N. and A.V. analysed the results and wrote the manuscript. All authors reviewed and revised the manuscript. 

\section*{Additional information}

\textbf{Accession codes}: Data is available upon reasonable request. \\
\textbf{Competing interests}: There are no competing interests. 

\section*{SI}
\label{sec:SI}

\subsection*{Animations of Evolution} To supplement the description for the evolution of smectic defect density distributions presented in Figures \ref{fig:grainDynamics} and \ref{fig:fig3}, we provide visualization animations which can be found at 
\begin{center}
\texttt{https://datashare.tu-dresden.de/s/jXJ5yBNg3wKZF94}    
\end{center}

\subparagraph*{EvolSmecticLines} This video corresponds to  \ref{fig:grainDynamics}, showing the rescaled density field $\psi$ in the front and back perspectives on a linear time scale.

\subparagraph*{D450\_V0.5} Here we combine the defect distributions along 100 logarithmic temporal bins to an animation. On the left side the transient of nematic distortion energy is depicted, while the red diamonds indicate the domain of the temporal bin. The right side provides the defect density distribution for the time bin. These results correspond to the snapshots presented in  \ref{fig:fig3} [A].

\subparagraph*{R100\_V0.5} Analog to \textit{D450\_V0.5}, this animation corresponds to the result presented in  \ref{fig:fig3} [B]. The defect densities have been derived from simulation results shown in \textit{EvolSmecticLines}.

\end{document}